\documentclass[%
aip,
sd,%
amsmath,amssymb,
reprint,%
]{revtex4-1}
\usepackage[colorlinks=true,
linkcolor=red,
urlcolor=black,
citecolor=blue]{hyperref}
\usepackage{graphicx}
\usepackage{epstopdf}
\usepackage{dcolumn}
\usepackage{bm}
\usepackage{subfigure}

\begin{document}
\preprint{AIP/123-QED}

\title{Spiral wave chimera-like transient dynamics in three-dimensional grid of diffusive ecological systems}

\author{Bidesh K. Bera}\email{bideshbera18@gmail.com}
\affiliation{Department of Solar Energy and Environmental Physics, BIDR, Ben-Gurion University of the Negev, Sede Boqer Campus, Midreshet Ben-Gurion 8499000, Israel}
\author{Srilena Kundu}\email{srilena93@gmail.com}
\affiliation{Physics and Applied Mathematics Unit, Indian Statistical Institute, 203 B. T. Road, Kolkata-700108, India}
\author{Paulsamy Muruganandam}\email{murganand@gmail.com}
\affiliation{Department of Physics, Bharathidasan University, Tiruchirapalli-620024, India}
\author{Dibakar Ghosh}\email{diba.ghosh@gmail.com}
\affiliation{Physics and Applied Mathematics Unit, Indian Statistical Institute, 203 B. T. Road, Kolkata-700108, India}
\author{M. Lakshmanan}\email{lakshman.cnld@gmail.com}
\affiliation{Department of Nonlinear Dynamics, School of Physics, Bharathidasan University, Tiruchirapalli-620024, India}

\date{\today}

\begin{abstract}
In the present article, we demonstrate the emergence and existence of the spiral wave chimera-like transient pattern in coupled ecological systems, composed of prey-predator patches, where the patches are connected in a three-dimensional medium through local diffusion. We explore the transition scenarios among the several collective dynamical behaviors together with transient spiral wave chimera-like states and investigate the long time behavior of these states. The transition from the transient spiral chimera-like pattern to the long time synchronized or desynchronized pattern appears through the deformation of the incoherent region of the spiral core. We discuss the transient dynamics under the influence of the species diffusion at different time instants. By calculating the instantaneous strength of incoherence of the populations, we estimate the duration of the transient dynamics characterized by the persistence of the chimera-like spatial coexistence of coherent and incoherent patterns over the spatial domain. We generalize our observations on the transient dynamics in three-dimensional grid of diffusive ecological systems by considering two different prey-predator systems.

\end{abstract}


\maketitle

\begin{quotation}
		{\bf Chimera state, characterized by the spatial coexistence of coherent and incoherent dynamics has been an intriguing research area in complex dynamical systems. Recently, the exploration on the emergence of fascinating spiral wave chimera state in spatially extended systems has been established by many researchers. However, they mostly focus on the long term stable dynamical patterns over the considered domain. But, in ecology, the persistence and coexistence of species are largely determined by the short term transient behavior rather than the long time dynamics which may lead to the extinction of species. Here, we explore the transient dynamics of locally diffusive ecological systems over a three-dimensional spatial domain which leads to the emergence of spiral wave chimera-like transient pattern. As time progresses, the transient chimera-like pattern evolves towards the coherent or incoherent spatial pattern through the deformation of the spiral core, depending on the species diffusion strengths. We estimate the transient duration from the initial chimera-like dynamics to the long time dynamics by calculating the strength of incoherence of the populations.}
		
\end{quotation}


\section{Introduction}\label{intro}

In the literature of nonlinear dynamics, the fascinating phenomenon of spatial coexistence of incoherent and coherent populations in a network of identically coupled oscillators has become a topic of great interest. 
Such emergent phenomenon was first observed by Kuramoto and Battogtokh in 2002 in nonlocally coupled phase oscillators \cite{kuramoto} and later this state was named as \textit{chimera state} \cite{strogatz} by Abrams and Strogatz. Since then, the existence of such states has been established in networks of limit-cycle oscillators \cite{limit,limit2}, chaotic oscillators \cite{chaotic}, neuron models \cite{hr_ijbc}, etc. Later on, such hybrid states have also been observed in networks composed of global \cite{global1,global2,global3,global4,global5,global6,global7} and local \cite{laing,hr_bera1,hr_bera2,local1,local_2d,chimera_2d,local_3d} couplings, as well as in networks of some unusual interaction topologies \cite{complex_ch1, ch_hetero, chimera_multiplex, multiplex1, multiplex2, multiplex3, multiplex4, solitary, complex_ch2, chimera_modular, chimera_multiscale, multiplex5}. The quantification and robustness of chimera states together with incoherent and coherent states have been investigated \cite{bs_chimera}. Apart from theoretical prescriptions, the evidence of chimera states have also been confirmed in some experimental investigations \cite{chem_chi,elchem_chi,elec_chi,opto_chi,mech_chi,feedback_chi}. The detection of such peculiar behavior from different aspects are presented \cite{chimera_rev,epl_rev,plr}.        

\par On the other hand, the interaction among the underlying dynamical units often gives rise to fascinating spatiotemporal patterns that have immense applicability in various fields of science, especially in biology and ecology. Among these emergent patterns, spiral waves \cite{sw1,sw2,sw3} are of ubiquitous importance which have been observed in a wide range of spatially extended systems of excitable or self-oscillatory elements. This universal class of spatiotemporal patterns have been experimentally verified in various chemical and biological systems \cite{sw_exp1}. Moreover, spiral waves are thought to be highly relevant to some pathological conditions in cardiac systems such as ventricular fibrillation \cite{app1,app2}, and to the neural context including human visual cortex \cite{app3}, hippocamal slices \cite{app4} and even in epileptic seizures \cite{app5}. 

\par Again, nonlocal interactions in spatially extended two-dimensional \cite{swch1,swch2,swch3,swch4} or three-dimensional \cite{swch5} systems can induce anomalous spiral dynamics (spiral wave chimeras) that possess phase locked oscillators in the spiral arms along with a core composed of phase randomized oscillators. This sort of phenomenon was first reported by Shima and Kuramoto \cite{swch1} for a paradigmatic three component reaction-diffusion system, in which two components were diffusion free, while the third component playing the role of coupling agent was eliminated adiabatically in order to generate the nonlocal effect. Later, Martens et al. \cite{swch2} provided the analytical description of such spiral wave chimeras in nonlocally coupled phase oscillators. Besides, they also calculated the speed of rotation and size of the incoherent core with the help of perturbation theory. Complex oscillatory and locally chaotic homogeneous systems in the presence of nonlocal coupling may also give rise to spiral chimera states \cite{swch3}. Further, spiral chimera waves were also observed in the three variable Hastings-Powell model \cite{swch4} containing a single diffusible variable under a set of dynamical conditions. Experimental verifications of nonlocally coupled Belousov-Zhabotinsky chemical oscillators \cite{swch6,swch7} have also been reported. Very recently, the emergence of spiral wave chimeras have been established in two-dimensional array of globally coupled phase oscillators in the presence of heterogeneous phase lags \cite{swch_global}. Apart from these, occurrence of such states in locally coupled three-component reaction-diffusion systems \cite{swch8} has also been reported. In addition, recently, spiral wave chimeras have been investigated in prey-predator metapopulation models \cite{swch9,swch_new} in the presence of local diffusion over a two-dimensional landscape.

Most studies in ecology consider species movement in two-dimensions. However, the real-world is three-dimensional and several marine species and birds usually migrate through volumetric environments \cite{eco_3dmovement1, eco_3dmovement2}. So, instead of considering surface movement in a two-dimensional plane, study of diffusion in three-dimensional space can properly imitate the movement of several species in ecosystems. 

Persistence and stability have been two of the long discussed topics of ecosystems in population ecology. Although, previously the ecological studies were mostly biased towards the long-term dynamics of these systems, the recent researches focus on their short term behavior. The long term dynamics of real ecological systems can lead to extinction, whereas in an ecologically relevant shorter timescale, the systems persist. This emphasizes the importance of transient dynamics \cite{eco_transient1, eco_transient2, eco_transient3} to understand the persistence and coexistence of ecological systems. In planktonic systems \cite{plankton1, plankton2}, exploration of transient dynamics enables one to determine the key factors leading to the coexistence of species in lakes during seasonality. Further, in explaining the distribution and abundance of species or in the study of insect and disease outbreaks \cite{epidemic}, transient dynamics has been a crucial area of research.

\par Motivated by the above, in the present article, we investigate the transient dynamics of coupled ecological prey-predator systems over a three-dimensional diffusive medium. Here we observe that the  transient dynamics is induced by local diffusion of species in 3D cubic medium of coupled ecological systems and interestingly a spatial spiral chimera-like pattern emerges in the transient period. Such pattern occurs through the deformation of the spiral core initiated from a coherent spiral wave. In our work, we have observed that the deformation of spiral wave not only depends on the diffusion coefficients, but also on the transient time. Although the transient period of individual prey-predator system is shorter, the collective dynamics of the coupled prey-predator system stimulates a long transient period, during which interesting spatiotemporal patterns are observed that play vital role in several ecological processes. Here we estimate the transient time for the transition from initial dynamics to the long term dynamics taking account of the effect of diffusion coefficients. Previously, transient chimera states \cite{transient_ch1, transient_ch2, transient_ch3} were reported by several researchers in mostly 1D and 2D domains for smaller system sizes. However, these studies mention a sudden collapse of transient chimera states and abrupt transition to coherent states as the time grows. In contrast to this, here we observe that the size of the deformed core of the spiral wave gradually increases as time grows and ultimately leads to the desynchronized or synchronized dynamics in the three-dimensional space depending on the diffusion coefficients for sufficiently large system size. To testify the existence of transient chimera-like dynamics and transition to incoherent or coherent dynamics, here we consider two different types of prey-predator systems with Holling type response functions \cite{holling}; one is the traditional Lotka-Volterra type system with Holling III functional response, and the second system models aquatic plankton dynamics (phytoplankton-zooplankton system) in the presence of Holling II functional response. We explore the transition scenario among various dynamical states mediated via spiral chimera states as the diffusion coefficients increase. All these states are characterized by calculating the strength of incoherence measure \cite{lakshman_measure}.


\par The subsequent parts of the paper are organized as follows. In the next section \ref{gmf}, we provide the general mathematical framework of the locally diffusive coupled ecological systems over a cubic domain. After that, we investigate the influence of diffusive coefficients and demonstrate the emergence of transient spiral wave chimera patterns together with other dynamical states as well in prey-predator systems coupled via local diffusion in three dimensional grids by considering two different ecological models. 
The results of Lotka-Volterra type system with Holling III interaction and the phytoplankton-zooplankton system with Holling II interaction are presented in sections \ref{pp1} and \ref{pp2}, respectively. Finally, we conclude our observations in section \ref{con}.

\section{General Mathematical Framework}\label{gmf}
 The general mathematical form of a prey-predator community can be described by a reaction-diffusion system \cite{lakshmanan_book},
\begin{equation}\label{eq1}
\begin{array}{lcl}
\frac{\partial X}{\partial t} = f(X,Y) + D_1 \nabla^2X,\\\\
\frac{\partial Y}{\partial t} = g(X,Y) + D_2 \nabla^2Y.
\end{array}
\end{equation}
Here $X$ and $Y$ correspond to the prey and predator population densities respectively. $D_1$ and $D_2$ are respective diffusion coefficients of prey and predator communities, which can be equal or unequal according to  the choice of particular system and interaction. The functions $f(X,Y)$ and $g(X,Y)$ can be formalized as \cite{model2}, $f(X,Y) = P(X) - E(X,Y)$, $g(X,Y) = \gamma E(X,Y) - \mu Y$, determined by the internal biological processes in the community. The function $P(X)$ describes the local growth and natural mortality of the prey species, whereas $E(X,Y)$ describes the functional response between prey and predator, i.e., prey mortality via predation. The parameter $\gamma$ is related to the number of newly incorporated predators to the community via predation and $\mu$ is the intrinsic mortality rate of the predator. The functions $P(X)$ and $E(X,Y)$ can be chosen accordingly depending on the type of prey population and the functional interaction of the predator.

\par
\begin{figure}[ht]
	\begin{center}
		\includegraphics[scale=0.25]{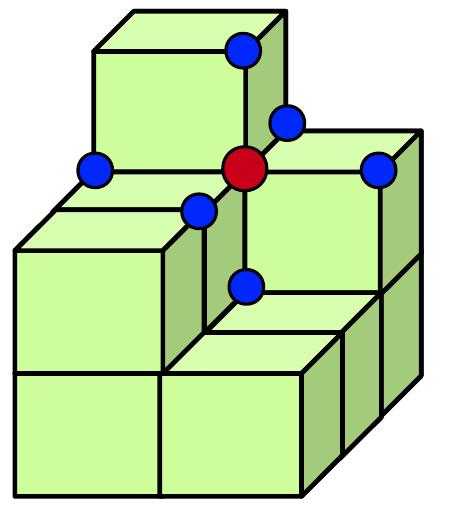}
	\caption{Schematic representation of a locally coupled network in three dimensional cubic lattice formation. Here red circle is connected to its six neighboring blue nodes, two nodes in each perpendicular directions. 
	\label{fig1}}
	\end{center}
\end{figure}

 Here we focus our study on the evolution of transient dynamical patterns in three spatial dimensions. In this case, the general mathematical form becomes,
\begin{equation}\label{eq2}
\begin{array}{lcl}
\frac{\partial X(x,y,z,t)}{\partial t} = f(X,Y) + D_1\big[\frac{\partial ^2 X}{\partial x^2} + \frac{\partial ^2 X}{\partial y^2} + \frac{\partial ^2 X}{\partial z^2}\big],\\\\
\frac{\partial Y(x,y,z,t)}{\partial t} = g(X,Y) + D_2\big[\frac{\partial ^2 Y}{\partial x^2} + \frac{\partial ^2 Y}{\partial y^2} + \frac{\partial ^2 Y}{\partial z^2}\big],
\end{array}
\end{equation}
for $0<x<L_x, 0<y<L_y, 0<z<L_z$, where $L_x, L_y, L_z$ respectively denote the length, breadth and height of the cubic domain. For numerical simulations, we discretize the domain on a $N\times N\times N$ cubic lattice with $N=100$ lattice points in each direction. The pictorial representation of the cubic lattice is depicted in Fig. \ref{fig1}, where each node (red circle) interacts with its six nearest neighbor nodes (blue circle) via diffusion. The numerical simulations are performed using Euler integration method with spatial step $h=1.0$ and integration time step $dt =0.01$, along with zero-flux boundary conditions. The initial conditions for diffusion over three-dimensional cubic domain are chosen as 
\begin{equation}
\begin{array}{lcl}
X_{i,j,k} = X^* - \epsilon_1 (j + k - N) ,\\\\
Y_{i,j,k} = Y^* - \epsilon_2 (i + k - N),
\end{array}
\end{equation}
where $(i,j,k)$ is the lattice index for $i,j,k = 1,2, ..., N$. $(X^*,Y^*)$ is the steady state of the system \eqref{eq2} in the absence of diffusion and $\epsilon_1, \epsilon_2$ are small fluctuations added to the steady state. Here we consider $\epsilon_1=\epsilon_2 = 10^{-4}$.

\section{Prey-Predator model}\label{pp1}

First, we explore the transient dynamics by considering the following functional forms \cite{model1},
\begin{equation}\label{eq4}
P(X) = X(a-X), ~~~ E(X,Y) = \frac{bX^2Y}{1+X^2},
\end{equation}
where the first function $P(X)$ signifies the logistic growth of prey species and the second form $E(X,Y)$ represents the Holling III functional response \cite{holling} of the predator. In the absence of diffusion, the system \eqref{eq2} along with \eqref{eq4} exhibits oscillatory dynamics for a fixed set of parameter values \cite{swch9,model1} $a = 6.46, b=1.25, \gamma = 0.64$ and $\mu = 0.5$. The system admits a stable limit cycle around the interior equilibrium point $(X^*,Y^*)$, where $X^*=\sqrt{\frac{\mu}{b\gamma - \mu}}$ and $Y^* = \frac{\gamma (a-X^*)}{X^*(b\gamma - \mu)}$. 

\begin{figure*}[ht]
 	\centerline{
 		\includegraphics[scale=0.42]{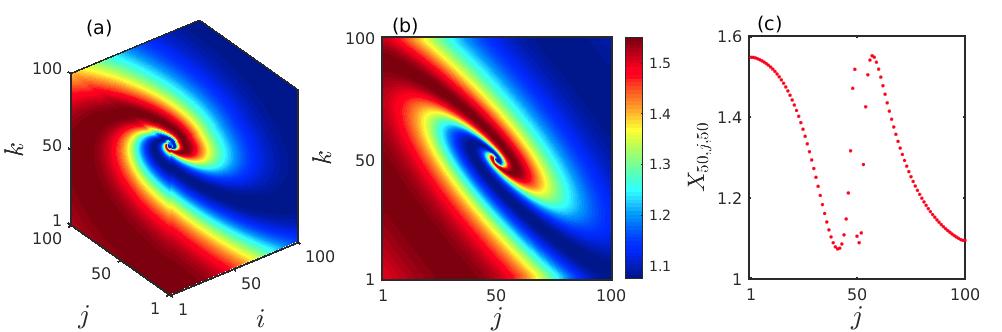}}
 	\caption{Snapshots of the spatial evolution of prey population $X$ depicting the spiral chimera-like spatial correlation among the prey sites with $N=100$ at a particular time instant $t=5000$ in (a) 3D domain, (b) 2D domain taking cross-section along $i=N/2$, and (c) 1D snapshot for fixed $i = k =N/2$. Diffusion coefficients are fixed at smaller values, $D_1 = D_2 = 0.0001$.}
 	\label{fig2}
 \end{figure*}
 
 \par Now we numerically investigate the different collective dynamical states together with transient chimera-like patterns in the 3D grid of the considered model. In presence of very low prey and predator diffusion rates $D_1=D_2=0.0001$, the evolution of spatial pattern by the prey populations $X_{i,j,k}$ at a particular time $t=5000$ is depicted in Fig. \ref{fig2}. The emergence of spiral patterns over the cubic domain, as well as over the two-dimensional plane by taking cross-section along $i = N/2$, are shown in Figs. \ref{fig2}(a) and \ref{fig2}(b), respectively. Further, to clearly demonstrate the spatial correlation among the prey-predator sites in an 1D array, we plot the prey abundances  $X_j$ in Fig. \ref{fig2}(c), taking a horizontal cross-section passing through the center of the core ($k=N/2$) of Fig. \ref{fig2}(b). These figures clearly demonstrate that the spatial incoherence in a small neighborhood along the center coexists with the spatially coherent surrounding oscillators, which eventually takes the form of a chimera state.
  
 \begin{figure*}
  \begin{subfigure}
    \centering\includegraphics[scale=0.42]{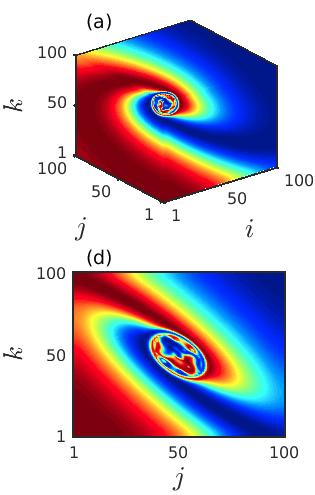}
  \end{subfigure}
  \begin{subfigure}
    \centering\includegraphics[scale=0.42]{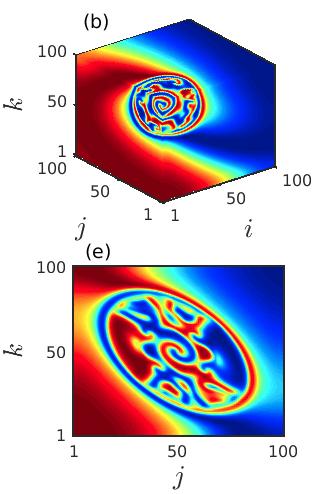}
      \end{subfigure}
  \begin{subfigure}
    \centering\includegraphics[scale=0.42]{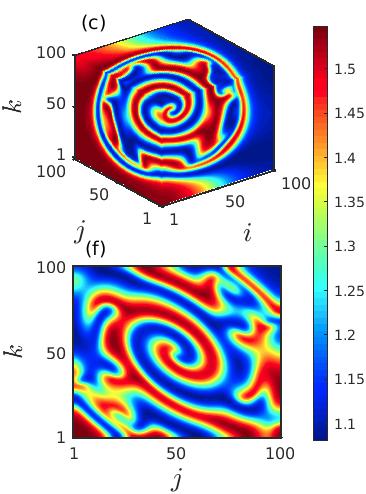}
  \end{subfigure}
      \caption{Snapshots of the spiral chimera-like pattern in the prey population $X$ at a particular instant $t=5000$ exhibiting the deformation of the spiral core in the three dimensional and two dimensional lattice with respect to three different combinations of the diffusion coefficients. The upper panel illustrates the patterns emerging in the cubic domain and the lower panel demonstrates the patterns over the 2D plane by taking cross-section along $i=N/2$. Diffusion coefficients are assigned the values (a, d) $D_1 = D_2 = 0.001$, (b, e) $D_1 = D_2 = 0.005$, and (c, f) $D_1 = D_2 = 0.02$. The colorbar represents the prey population density $X_{i,j,k}$ over the spatial domain.}
       	\label{fig3}
\end{figure*}

 \par Next, we explore the deformation of the spiral core in the cubic lattice of ecological systems by changing the diffusion coefficients $D_1$ and $D_2$. Figure \ref{fig3} portrays the snapshots of the prey population $X$ at the particular time instant $t=5000$, demonstrating the instability of the spiral core which leads to the coexistence of spatial coherence and incoherence as the diffusion strengths are varied. The appearance of the spiral chimera-like dynamics is noticed in the cubic domain as well as in the 2D plane by taking the cross-section along $i=N/2$ plotted respectively in the upper and lower panels. For lower diffusion coefficients $D_1=D_2=0.001$, the initiation of the spiral chimera-like state is shown in the left column of Fig. \ref{fig3}. A smaller region in the spiral core area exhibiting spatial incoherence is surrounded by the spatially coherent populations. The typical snapshots in 3D domain and 2D plane (slicing along $i=N/2$) are plotted in Figs. \ref{fig3}(a) and \ref{fig3}(d), respectively. As the diffusion rates of the prey and predator populations increase, the destruction of the spiral at the center starts to increase and hence the size of the spiral core enlarges. For slightly increased values of the diffusion strengths at $D_1=D_2=0.005$, the formation of the irregular patterns along the center of the spiral are depicted in Figs. \ref{fig3}(b), \ref{fig3}(e), respectively. Although, interestingly, here the origination of another spiral wave from the core area is noticed. Consequently, for further increase in the diffusion rates, the region depicting the irregular patterns surrounding the newly emerging spiral wave enlarges. In fact, the spiral wave that starts from the center also develops with the enhancement of the diffusion coefficients. This phenomena is depicted in the Figs. \ref{fig3}(c), \ref{fig3}(f), for $D_1=D_2=0.02$.

\par
\begin{figure*}
  \begin{subfigure}
    \centering\includegraphics[scale=0.42]{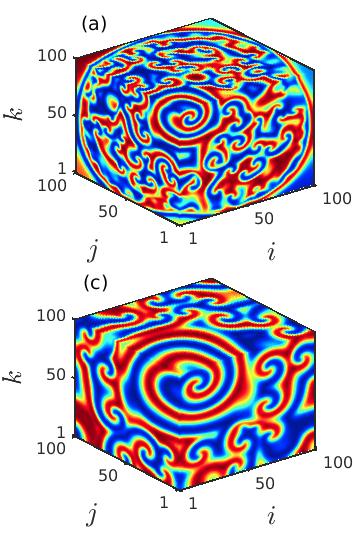}
  \end{subfigure}
  \begin{subfigure}
    \centering\includegraphics[scale=0.42]{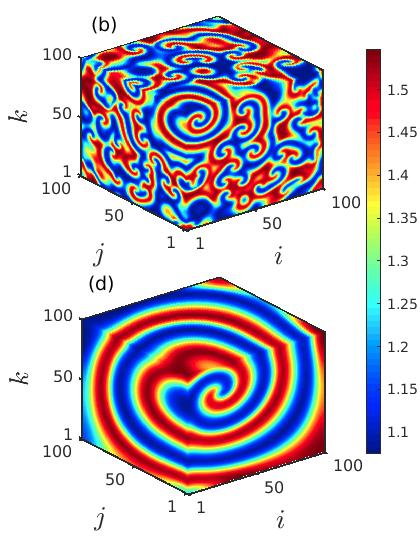}
      \end{subfigure}
      \caption{Snapshots of the spatial distributions of the prey species $X_{i,j,k}$ over the cubic domain  at a particular time instant $t=10000$ demonstrating the development of the newly emerging spiral originating from the core with respect to different choices of the diffusion coefficients. Diffusion strengths are fixed at (a) $D_1 = D_2 = 0.008$, (b) $D_1 = D_2 = 0.01$, (c) $D_1 = D_2 = 0.02$, and (d) $D_1 = D_2 = 0.05$.}
       	\label{fig4}
\end{figure*}

 Further, in Fig. \ref{fig4}, we demonstrate the development of the newly emerging spiral wave originating from the core at a much higher time instant $t=10000$. Four different sets of diffusion coefficients $D_1= D_2 = 0.008, 0.01, 0.02$, and $0.05$ are considered in Figs. \ref{fig4}(a), \ref{fig4}(b), \ref{fig4}(c) and \ref{fig4}(d), respectively, to display that with the increase of the diffusion strengths the size of the incoherent core increases. Inside the incoherent core the newly emerging spiral gradually develops which is surrounded by multiple tiny spirals throughout the incoherent core region. As the diffusion strengths increase further the number of tiny spirals reduces and the spiral emerging from the center spreads all over the spatial domain and eventually one giant spiral is noticed in the cubic domain (cf. Fig. \ref{fig4}(d)). 

 \begin{figure}[ht]
 	\centerline{
 	\includegraphics[scale=0.36]{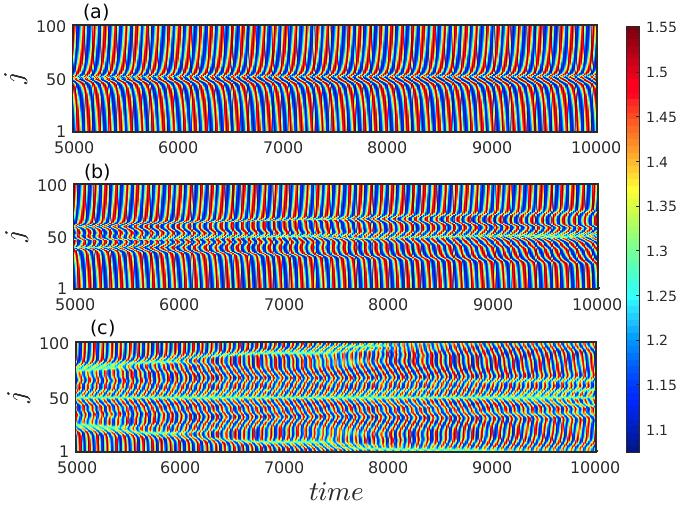}}
 	\caption{Spatiotemporal evolution of the prey population $X_{i,j,k}$ along the cross section $i = k = N/2$ highlighting the stationarity of the observed spiral dynamics for three distinct sets of diffusion coefficients (a) $D_1 = D_2 = 0.0001$, (b) $D_1 = D_2 = 0.001$, and (c) $D_1 = D_2 = 0.005$, respectively. The colorbar represents the prey population density.}
 	\label{fig5}
 \end{figure}
 
 \par To investigate the stationarity of the observed spiral dynamics, we plot the spatio-temporal evolution of the prey populations $X_{i,j,k}$ in Fig. \ref{fig5} by taking cross-section along $i=k=N/2$. Figure \ref{fig5}(a) represents the space-time behavior of the initial spiral chimera-like pattern as depicted in Fig. \ref{fig2} for diffusion coefficients $D_1= D_2 =0.0001$. The long time evolution during the time interval $[5000, 10000]$ exhibits the pattern to be stationary as there is no such variation of the pattern with respect to time. Irregularity among few members of the populations along the core is noticed from the figure. From the spatiotemporal evolution it is also evident that the spiral initiating from the core rotates along the outward direction. However, the space-time plots in Figs. \ref{fig5}(b, c) for diffusion coefficients $D_1 = D_2 = 0.001$ and $D_1=D_2=0.005$ display that the regions showing irregular motion along the center continue to enlarge as time grows. This observation suggests that the chimera-like spatial coexistence of regular and irregular motion observed in the cubic domain is not stable with respect to time and thereby represents the transient dynamics. 
 
  \begin{figure*}
  \begin{subfigure}
    \centering\includegraphics[scale=0.42]{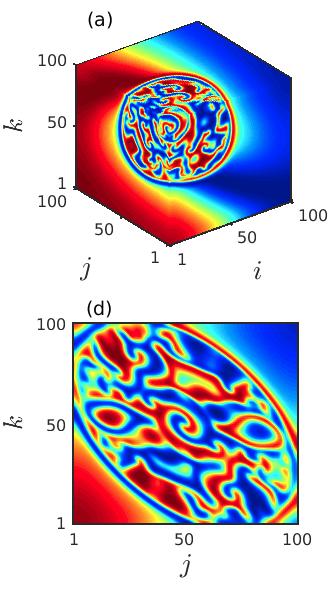}
  \end{subfigure}
  \begin{subfigure}
    \centering\includegraphics[scale=0.42]{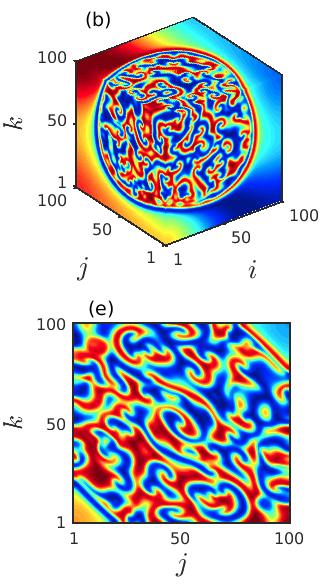}
      \end{subfigure}
  \begin{subfigure}
    \centering\includegraphics[scale=0.42]{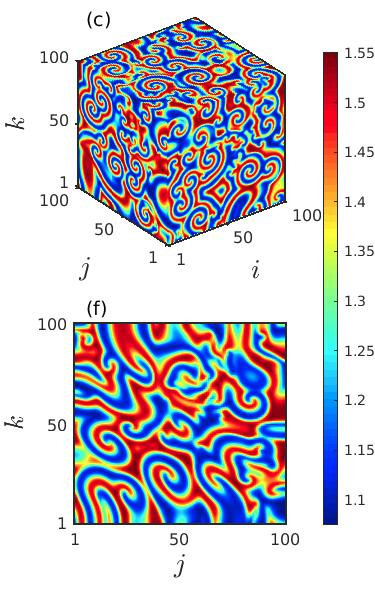}
  \end{subfigure}
      \caption{Spatial distributions of the prey species $X_{i,j,k}$ for fixed diffusion strengths $D_1 = D_2 = 0.005$ at three distinct time instants (a, d) $t=7000$, (b, e) $t=10000$, and (c, f) $t=30000$, respectively, which substantiate the transient behavior of the observed pattern. The upper panel shows the snapshots of the patterns emerging in the cubic domain and the lower panel corresponds to the snapshots in the 2D plane by taking cross-section along $i=N/2$. }
       	\label{fig6}
\end{figure*}

\par In order to explore the transient behavior of the considered prey-predator system over the cubic domain, we keep track of the variation in the spatial patterns with the advancement of time. The spatial distribution of the prey species $X_{i,j,k}$ at three particular time instants $t=7000$, $t=10000$ and $t=30000$ are delineated in the left, middle and right columns of Fig. \ref{fig6} for fixed diffusion coefficients $D_1 = D_2 = 0.005$. The figure manifests that as the time grows the size of the spiral core with irregular patterns increases and the region with coherent dynamics decreases. Beyond a particular time instant the coherent region vanishes completely and incoherent patterns prevail in the entire 3D domain. These occurrences are clear from both the 3D and the 2D spatial dynamics plotted in the upper and the lower panels. Also, from the figures it can be perceived that the destruction of the spiral core gives rise to multiple small spiral patterns in the entire cubic domain.

\begin{figure*}[ht]
	\centerline{
		\includegraphics[scale=0.46]{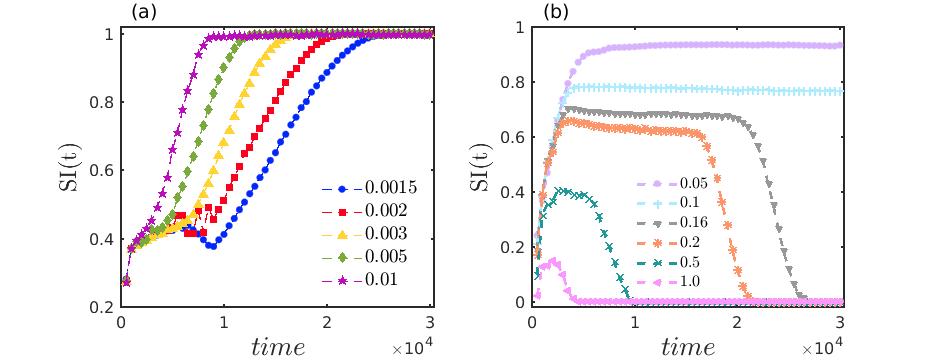}}
	\caption{The variation in the transient duration characterized through local SI(t) measurement. Temporal evolution of SI(t) with respect to time for different values of diffusion coefficients $D_{1,2}$. (a) Lower values of $D_{1, 2}$ eventually lead to the incoherent dynamics in the long time, (b) beyond a certain range of $D_{1,2}$, the dynamics shifts from incoherent to coherent dynamics with the advancement of time. The values indicated in the legends correspond to the cases of identical diffusion coefficients $D_{1,2}$. Here the volume of the cubic bin $n=4^3$ and threshold in Eq. \eqref{eq7} $\delta_1 = 0.01$ are considered.}
	
	\label{fig7}
\end{figure*}

\par Now, we investigate the temporal features of the transient spiral chimera-like dynamics by using the notion of strength of incoherence (SI) measure. Here we characterize the chimera-like patterns together with other dynamical states by using the quantitative statistical SI measurement as developed by Gopal {\it et al.} \cite{lakshman_measure}. Our observed spiral chimera-like dynamics differs from the stationary chimera states as it emerges during the transient period. So, to examine the temporal variations of the spatial patterns, we calculate the instantaneous SI as a function of time. For this purpose, first we define a difference variable $w_{i,j,k}$ corresponding to the state variable $X_{i,j,k}$ as
	\begin{equation}
		\begin{array}{lcl}
		w_{i,j,k} = \Big[(X_{i,j,k}-X_{i+1,j,k})^2 + (X_{i,j,k}-X_{i,j+1,k})^2 \\~~~~~~~~~~~+ (X_{i,j,k}-X_{i,j,k+1})^2 \Big] ^\frac{1}{2},~~ i,j,k = 1, 2, ..., N.
		\end{array}
	\end{equation}
	Now we divide the entire cubic lattice into $M(=M_h \times M_w \times M_b)$ number of cubic bins each of equal volume $n(=N/M_h \times N/M_w \times N/M_b)$ and then we calculate the local standard deviation $\sigma(p,q,r)$ in each of these bins as
		\begin{equation}
		\begin{array}{lcl}
		\sigma(p,q,r)(t) \\\\ = \sqrt{\frac{1}{n^3}\sum\limits_{i=n(p-1)+1}^{np}~\sum\limits_{j=n(q-1)+1}^{nq}~\sum\limits_{k=n(r-1)+1}^{nr}(w_{i,j,k}-\langle w \rangle)^2},
		\end{array}
		\end{equation}
		where $\langle w \rangle = \frac{1}{N^3} \sum\limits_{i=1}^{N}\sum\limits_{j=1}^{N}\sum\limits_{k=1}^{N} w_{i,j,k}$; $p=1,2,...,M_h; q=1,2,...,M_w$; and $r=1,2,...,M_b$. Then the instantaneous strength of incoherence is defined as 
		\begin{equation}\label{eq7}
		\begin{array}{lcl}
		\mbox{SI(t)} = 1-\frac{\sum\limits_{p=1}^{M_h}\sum\limits_{q=1}^{M_w}\sum\limits_{r=1}^{M_b}s{(p,q,r)}(t)}{M},
		\end{array}
		\end{equation}
	where $s{(p,q,r)}(t) = \Theta(\delta_1-\sigma(p,q,r)(t))$. Here $\Theta(.)$ is the Heaviside step function and $\delta_1$ is the predefined threshold value \cite{delta1, delta2}. Consequently, the values $\mbox{SI(t)}=1$ and $\mbox{SI(t)}=0$ characterize the incoherent or desynchronized and coherent or synchronized dynamical patterns, respectively, at any instant of time. While for the spiral chimera-like spatial pattern observed here, $\mbox{SI(t)}$ takes value from the interval $(0, 1)$. Additionally, for the giant spiral pattern as observed in Fig. \ref{fig4}(d), due to the periodic fluctuations between the maximum and minimum species densities, SI(t) lies between 0 and 1.

Considering several sets of values of the diffusion coefficients $D_{1,2}$, the variation of instantaneous SI(t) is illustrated in Fig. \ref{fig7} with respect to time. The transition scenarios during the transient period  initiating from chimera-like state to completely incoherent states are shown in Fig. \ref{fig7}(a) for comparatively smaller diffusion rates. In this figure, $\mbox{SI(t)} =1$ corresponds to the emergence of completely incoherent state with multiple tiny spirals as depicted in the previous figures, whereas $\mbox{SI(t)} <1$ characterizes the development of spiral chimera-like state with incoherent core. For very low values of diffusion coefficients $D_{1,2}=0.0015$, the transition from spiral chimera-like state to the fully incoherent state through the enlargement of the core appears at the time instant $t=2.5\times 10^{4}$, represented by blue filled circles in Fig. \ref{fig7}(a).  Interestingly, with increasing values of the diffusion coefficients $D_1$ and $D_2$, the required time to reach the unit value by the instantaneous SI(t) decreases. Thus by considering several illustrative values of $D_{1,2}$, and using the corresponding values of the instantaneous SI(t), we estimate the transient duration of the spiral chimera-like states. It is observed that due to the increase of the diffusion strengths up to a certain range in the coupled system, the transient period depicting the chimera-like pattern is reduced, and consequently, the duration of the incoherent pattern is enhanced. However, beyond a certain range of the diffusion coefficients, as $D_{1,2}$ increases, the density of the tiny spirals in the incoherent region starts to reduce and the temporal evolution eventually converges to one giant spiral or to the completely synchronized dynamics with the enhancement of time and diffusion strengths. This phenomena is clearly depicted in Fig. \ref{fig7}(b) taking several values of the diffusion coefficients $D_{1,2}$. As $D_{1,2}$ increases the value of SI(t) starts to decrease from unit value and with the advancement of time and higher diffusion rates, SI(t) gives zero value which characterizes the emergence of completely coherent or synchronized dynamics. Non-zero and non-unit values of SI(t) in Fig. \ref{fig7}(b) indicate the emergence of one giant spiral in the entire cubic domain for which the spatial pattern is demonstrated in Fig. \ref{fig4}(d). With the increase of $D_{1,2}$, the frequency of rotation of the spiral wave gradually decreases and consequently, the values of SI(t) also reduce and tend to become zero. As the diffusion rates become higher, the dynamics rapidly converges towards the synchronized motion (spatial pattern not shown here) with much lesser transient duration. The transition scenarios between the transient and the long term dynamics are particularly depicted for the identical diffusion strengths of both the prey and the predator species. However, a similar phenomenon is noticed even if the diffusion coefficients of the prey species are smaller than that of the predator species, i.e., when the diffusion rates of the two species differ in their magnitudes (figure not shown).

\begin{figure}[ht]
	\centerline{
		\includegraphics[scale=0.425]{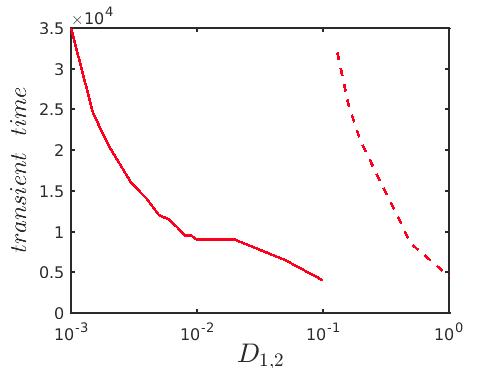}}
	\caption{Duration of the transient time beyond which a stable ecological dynamics is observed, with respect to the diffusion coefficients $D_{1,2}$. Solid red line demarcates the transition border from spiral chimera-like state to incoherent state, whereas the dashed line represents the transient duration before converging to the synchronized state.}
	\label{fig8}
\end{figure}

Subsequently, for better perception, we calculate the transient duration beyond which a stable ecological dynamics persists. Based on the instantaneous SI(t) measurement, the transient periods demonstrate the spatial coexistence of coherence and incoherence, which are estimated in Fig. \ref{fig8} against the diffusion coefficients $D_{1,2}$. The red solid line in Fig.~\ref{fig8} demarcates the transition from chimera-like state to incoherent state. From this plot it is clear that for lower diffusion coefficient values, one needs longer time to reach the incoherent motion, while for higher diffusion strengths the transient period corresponding to the spiral chimera-like state monotonically decreases. As observed previously in Fig. \ref{fig7} that beyond certain range of the diffusion coefficients $D_{1,2}$, the long term dynamics changes from incoherent to synchronized behavior, for which the variation in the transient duration is depicted by the red dashed plot in Fig. \ref{fig8}. So, from the figure, one can conclude that a strong diffusion in the locally coupled diffusive prey-predator system leads to a shrinkage of the transient duration and correspondingly speeds up the emergence of the stable long term dynamics.

\section{Phytoplankton-zooplankton model}\label{pp2}
Next we consider another type of ecological model to establish the universality of occurrence of the spiral chimera-like features in the cubic lattice of diffusive ecological systems. We explore the model which imitates the spatiotemporal dynamics of an aquatic community, particularly, a phytoplankton-zooplankton system, constituting the following functional forms \cite{model2}  
\begin{equation}\label{eq8}
P(X) = X(1-X), ~~~ E(X,Y) = \frac{XY}{X+\beta}.
\end{equation}
Here again the first term signifies the logistic growth of the prey species, but unlike the previous model, the second term here represents the Holling II functional response \cite{holling} of the predator. The constant $\beta$ is the half-saturation abundance of the prey species. As obtained from the linear stability analysis \cite{model2}, whenever $\beta < \frac{\gamma - \mu}{\gamma + \mu}$, the system \eqref{eq2} along with \eqref{eq8} possesses an unstable steady state $(X^*, Y^*)$, where $X^* = \frac{\mu \beta}{\gamma - \mu}$ and $Y^* = (1-X^*)(X^*+\beta)$, without the presence of the diffusion terms. In this case, the steady state remains surrounded by a stable limit cycle, for which the system exhibits oscillatory dynamics. Here the parameter values are considered as $\beta = 0.45, \gamma = 2, \mu = 0.6$ for numerical simulations.
  \begin{figure*}
  \begin{subfigure}
    \centering\includegraphics[scale=0.42]{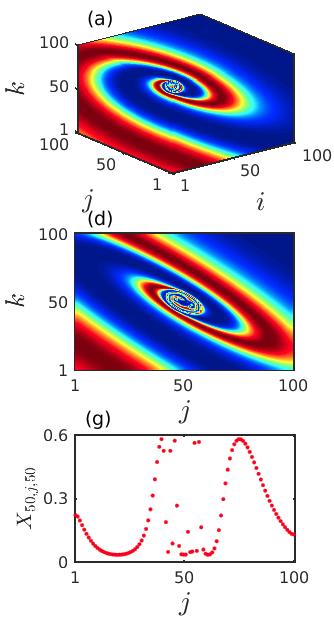}
  \end{subfigure}
  \begin{subfigure}
    \centering\includegraphics[scale=0.42]{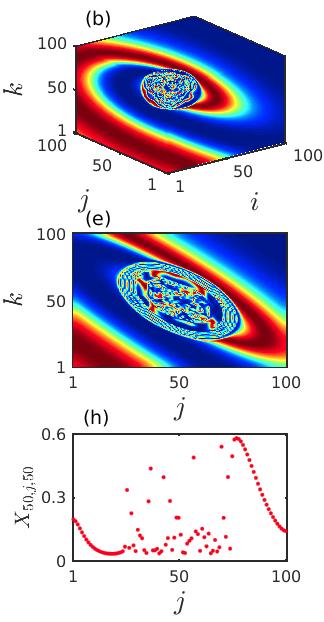}
      \end{subfigure}
  \begin{subfigure}
    \centering\includegraphics[scale=0.42]{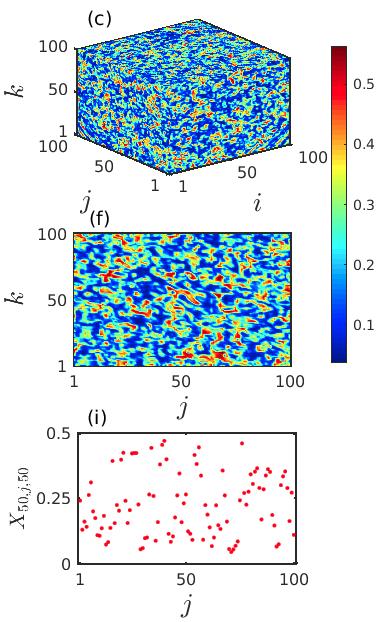}
  \end{subfigure}
      \caption{Snapshots of the spatial distributions of the population $X$ at a particular instant $t=5000$. The upper, middle and lower panels illustrate the patterns emerging in the cubic domain, over the 2D plane by taking cross-section along $i=N/2$, and in 1D array by taking cross-section along $i=k=N/2$. For the diffusion coefficients (a, d, g) $D_1 = D_2 = 0.0001$, spiral-chimera like pattern emerges with a small incoherent region at the core, (b, e, h) $D_1 = D_2 = 0.0005$, the incoherent core area gets enlarged, and (c, f, i) $D_1 = D_2 = 0.005$, the entire cubic domain exhibits incoherent dynamics. The colorbar represents the prey population density $X_{i,j,k}$ over the spatial domain.}
       	\label{fig9}
\end{figure*}

\par We integrate the system \eqref{eq2} along with \eqref{eq8} and observe spiral wave chimera-like states together with completely incoherent states. Similar to the previous case, here also we investigate the emergence and existence of chimera-like states during the transient period in local diffusively coupled ecological systems over the cubic domain. In Fig. \ref{fig9}, the particular time snapshots ($t=5000$) of spatial arrangements of the prey species (phytoplankton) are plotted where the color bar denotes the variation of the prey densities $X_{i,j,k}$. The upper, middle and lower panels represent the snapshots over the 3D grid, 2D plane by taking cross-section along $i=N/2$ and 1D array by taking cross-section along $i=k=N/2$, respectively. The initial distortion of coherent spiral core and initiation of the chimera-like state is observed for smaller diffusion strengths $D_1 = D_2 =0.0001$. The corresponding results are shown in the Figs.~\ref{fig9}(a, d, g). Here a very small incoherent region in the core area over the cubic domain is surrounded by the coherent populations. The incoherent portion of the chimera-like state gets enlarged with an increase in the diffusion coefficients. The spreading of such incoherent domain is demonstrated in the middle column of Fig.~\ref{fig9} for $D_1= D_2=0.0005$. For sufficiently strong diffusion coefficients $D_1=D_2=0.005$, the entire ecological network collectively produces incoherent dynamics. This happens due to the continuous expansion of the spiral core consisting of the incoherent populations with respect to the diffusion strengths. The corresponding time snapshots are shown in the right column of Fig.~\ref{fig9}. In the one dimensional snapshot presented in Fig.~\ref{fig9}(i), the scattered distribution of the state variables is bearing a strong resemblance of the existence of the incoherent states. Similar to the previous case, we can conclude that the strong diffusion among the different patches of the ecological network induces the transition from spiral chimera-like state to completely incoherent state through the continuous deformation of the incoherent core.

\begin{figure}[ht]
	\centerline{
		\includegraphics[scale=0.455]{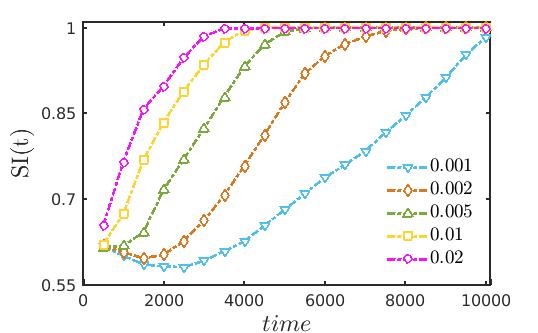}}
	\caption{ Variation of the instantaneous strength of incoherence SI(t) of the populations with respect to time depicting the transient duration before converging to the long time behavior for several choices of the diffusion coefficients $D_{1,2}$. The values indicated in the legend correspond to the case of identical diffusion coefficients $D_{1,2}$. Here also we consider $n=4^3$ and $\delta_1 = 0.01$.}
	\label{fig10}
\end{figure}

\begin{figure}[ht]
	\centerline{
		\includegraphics[scale=0.4]{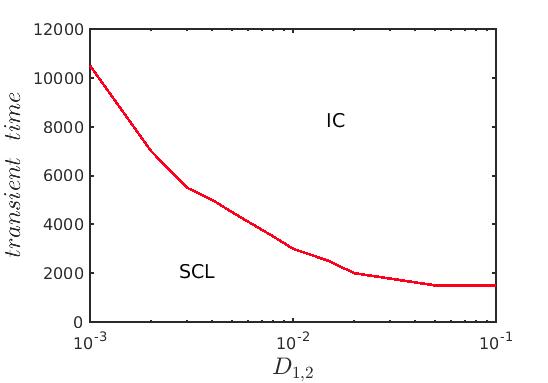}}
	\caption{Duration of the transient time until when the spiral chimera-like state persists with respect to the diffusion coefficients $D_{1,2}$. Solid red line demarcates the transition border from spiral chimera-like (SCL) transient pattern to long time incoherent (IC) dynamics.}
	\label{fig11}
\end{figure}

\par Similar to the case of previous model, we explore the transition scenario from transient chimera-like state to completely incoherent long term dynamics for several choices of diffusion strengths $D_{1,2}$. We compute the strength of incoherence SI(t) as a function of time $t$ using the formula \eqref{eq7} and the variations in the transient duration are plotted in Fig. \ref{fig10}. For each set of diffusion strengths, as time increases the evolution of the chimera-like state with the expansion of the core for $\mbox{SI(t)}<1$  is depicted in the figure. Finally, after a certain time the transient dynamics collapses and the long term incoherent dynamics is observed corresponding to the value $\mbox{SI(t)} =1$. Lastly, in Fig. \ref{fig11} for a better understanding of the variation of the transient duration with respect to the diffusion coefficients, we have plotted the time duration until when the chimera-like transient state persists against the equal diffusion coefficients $D_1= D_2$. The red solid line distinguishes the area of the chimera-like state from that of the incoherent state as delineated previously in Fig. \ref{fig8}. From the figure it is obvious that as the diffusion strengths $D_{1,2}$ increase the duration of the transient dynamics gradually decreases before converging to the long time incoherent behavior.

\section{Conclusion}\label{con}
In this paper, we have studied the emergence and existence of spiral wave chimera-like transient dynamics together with the long term incoherent and coherent spiral dynamics in locally diffusively coupled ecological systems. We considered the dynamics such that the entire system is interacting through diffusion in three spatial dimensions. We have focused our study on two ecological models, where one is the traditional Lotka-Volterra system and the other is the phytoplankton-zooplankton system, to establish the generalized occurrence of transient spiral chimera-like features in the locally coupled ecological systems in three dimensions. From precise spatiotemporal behavior, we can understand that the chimera-like coexistence of spatially coherent and incoherent dynamics over the cubic domain represents the transient phenomena. Using the instantaneous strength of incoherence measurement we have characterized the existence of various spatial patterns in short time duration as well as in long time duration. The transition among different dynamical states are portrayed here taking various diffusion strengths and at different time instants. At a particular time, the transition from spiral chimera-like pattern to incoherent pattern appears through the expansion of the incoherent core of the spiral chimera-like state with increasing diffusion strengths. Additionally, we have found that sufficiently large time may also induce the core distortion of spiral motion, with proper modulation of the diffusion coefficients. By measuring the instantaneous strength of incoherence, we analyzed the temporal features of the observed chimera-like patterns and estimated the transient time required corresponding to the complete spiral deformation, for which all the patches eventually reach the long time stable dynamics. Also it is observed that strong diffusion among the patches leads to the reduction of spiral deformation time.

\medskip

\par {\bf Acknowledgments:}
The work of P.M. is supported by DST-FIST (Department of Physics), DST-PURSE and MHRD RUSA 2.0 (Physical Sciences) Programmes. M.L. is supported by a Science and Engineering Research Board Distinguished Fellowship.

\medskip

\section*{Data availability}
The data that support the findings of this study are available within the article.

\end{document}